# Electric Polarization in ErCrO3 Induced by Restricted Polar Domains

V. A. Sanina a, *, B. Kh. Khannanov a, E. I. Golovenchits a, and M. P. Shcheglov a

a Ioffe Institute, St. Petersburg, 194021 Russia

*e-mail: sanina@mail.ioffe.ru

Abstract—Electric polarization in ErCrO3 single crystals has been investigated in the temperature range of 5–370 K. Ferroelectric ordering has not been found in any of the directions. However, electric polarization induced by restricted polar domains of structural origin has been observed. These domains are formed in the crystal matrix near impurity Bi3+ ions partially substituting Er3+ ions during the growth of single crystals by the method of spontaneous crystallization using solvent Bi$_2$O$_3$. The restricted polar domains form the superparaelectric state. Hysteresis loops with remanent polarization, both along the c axis and in the [110] directions, have been observed below some temperatures T$_{fr}$ (in the frozen superparaelectric state). The polarization exists up to certain temperatures, which depend on the applied electric field orientation with respect to the crystal axes and exceed significantly temperature T$_N$ of magnetic ordering. These temperatures correspond to the condition kT$_{fr}$≈ E$_A$ for activation barriers at the boundaries of the restricted polar domains.

## 1. INTRODUCTION

The polar order below the Néel temperature T$_N$ of magnetic ordering was previously found in orthochromites RCrO$_3$ with magnetic rare-earth R ions [1]. Temperatures T$_N$ for RCrO$_3$ are rather high (130–250 K) and significantly exceed the Curie temperatures T$_C$ of ferroelectric ordering of the previously studied multiferroics of the second type, in which the ferroelectric ordering is induced by magnetic ordering of a special type [2–7]. Somewhat later, the polar order was found in LuCrO$_3$ with nonmagnetic R ions below temperature T$_C$ exceeding T$_N$ [8].

Orthochromites RCrO$_3$ have an orthorhombically distorted perovskite structure with the centrosymmetric sp. gr. Pbnm [9, 10] forbidding ferroelectric ordering. It was suggested in [1] that the observed polar order in RCrO$_3$, occurring in preliminary applied electric field E, is fixed by the R–Cr exchange interaction below T$_N$. An RCrO$_3$ unit cell contains four formula units. Cr$^{3+}$ ions are located in oxygen octahedra. Axes of these octahedra deviate from the c axis, along which they are oriented in undistorted cubic perovskites. R$^{3+}$ ions are in strongly distorted polyhedra with eight nearest oxygen ions. The symmetry of local sites of R$^{3+}$ ions (Cs) is off-center; therefore, RO$_8$ quasi-molecules have nonzero electric dipole moments located in the (001) planes. However, these dipole moments in a unit cell deviate in different directions, thus compensating each other. The antiferroelectric ordering arises, which is consistent with the sp. gr. Pbnm.

The magnetic properties and magnetic phase transitions in RCrO$_3$ were analyzed in detail previously [9, 10]. Figure 1 shows magnetic structures occurring in RCrO$_3$. The antiferromagnetic position of spins along different crystal axes is characterized by vectors A, C, and G (G>> A, C). The weak ferromagnetic moment is described by vector F. In ErCO$_3$, the antiferromagnetic structure Γ1 (AxGyCz) is implemented at temperatures T ≤ $T_{N1}$= 5 K, while the structure Γ4 (GxAyFz) occurs at 5 K ≤ T ≤ $T_{N2}$= 130 K (the designations of the structures are given according to Bertaut [9]). In the temperature range of interest to us (T> 5 K), there are antiferromagnetic ordering Gx of spins of $Cr^{3+}$ ions along the x (a) axis and weak ferromagnetic moment Fz along the z(c) axis in $ErCrO_3$ (ECO).

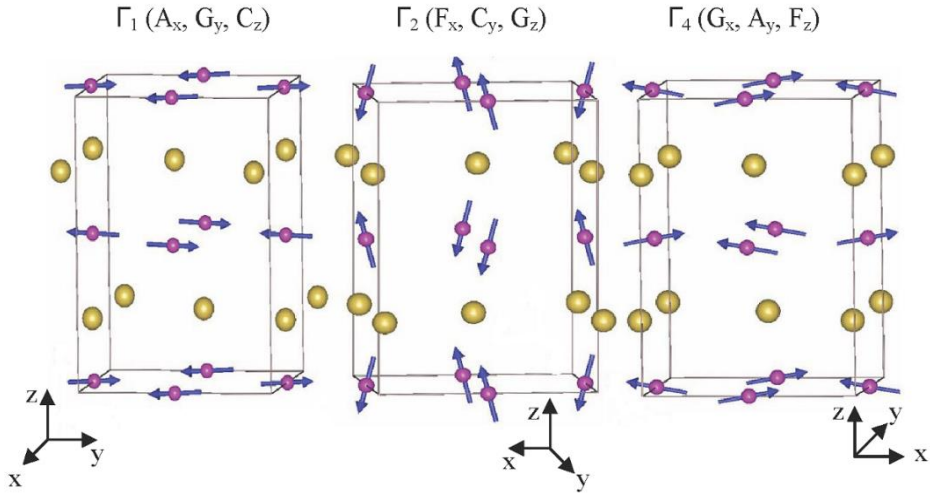

**Fig. 1.** Possible types of $RCrO_3$ magnetic structures. The arrows indicate spin orientations of $Cr^{3+}$ ions. The larger circles show the positions of $R^{3+}$ ions. The structural types are given in the Bertaut notation [9].

Figure 2 shows the temperature dependence of the dynamic magnetic susceptibility along the c axis. It can be seen that there is a well-defined magnetic phase transition near T = 130 K (referred to as $T_N$ below).

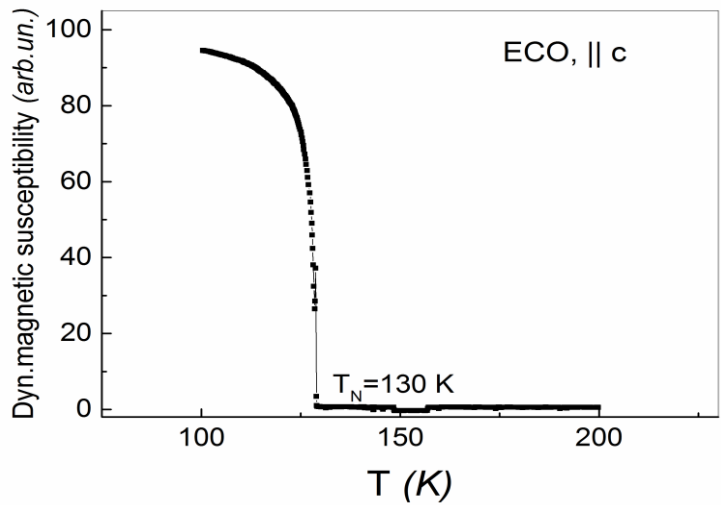

**Fig. 2.** Temperature dependence of the dynamic magnetic susceptibility for ECO along the *c* axis (for the *Fz* component of the weak ferromagnetic moment) at a frequency of 10 kHz.

$Er^{3+}$ ions (the ground state is $^4I_{5/2}$) have a large magnetic moment of ~9 µB with a large orbital contribution. Magnetic moments of $Er^{3+}$ ions are rigidly oriented along the c axis by the strong single-ion anisotropy due to the spin–orbit interaction, which significantly exceeds the crystal field [10]. …….Note that ceramic $RCrO_3$ samples generally having a large number of defects were investigated in [1, 8]. In this context, the obtained results should be checked on single crystals. In [11], we analyzed polar properties of $YCrO_3$ single crystals with a nonmagnetic R ion. Ferroelectric ordering was not found in any of the crystal directions in the temperature range of 5–400 K. However, electric polarization induced by restricted polar domains of both the magnetic and structural origin was observed. These domains were formed near $Pb^{4+}$ and $Pb^{2+}$ ions incorporated into the sites of $Y^{3+}$ ions during the growth of single crystals by the method of spontaneous crystallization in solvent $PbO + PbO_2$ [11]. The sizes of polar domains and correlation radii of electric dipole interactions were sufficient for the formation of restricted domains with ferroelectric ordering and activation barriers $E_A$ at their boundaries. The frozen superparaelectric state occurred at temperatures $kT_{fr} < E_A$, in which maxima of the pyroelectric current were observed near $T_{fr}$ and hysteresis loops of electric polarization were found at $T \leq T_{fr}$ [11]. These observations imitated a bulk-uniform ferroelectric phase transition.

The superparaelectric state of individual ferroelectric nanoregions (in the form of spheres) in a centrosymmetric dielectric matrix was theoretically studied in [12]. It was shown that the frozen superparaelec-tric state arises in a sample at the validity of certain conditions; the response of this state to the applied electric field E has the form of hysteresis loops with remanent polarization. These conditions are as follows. The restricted-domain sizes R should be smaller than the correlation radius Rc of interaction between electric dipoles but larger than the critical radius Rcr which allows the occurrence of ferroelectric order in the domain. At the validity of these conditions, all dipoles in the polar domains are oriented parallel. The surface screening of depolarization fields makes the formation of single-domain state of restricted polar domains favorable. The frozen superparaelectric state, at which the spontaneous repolarization of individual polar domains is impossible, arises at sufficiently low temperatures ($kT_{fr} \leq E_A$). This repolarization occurs at $kT \geq kT_{fr}$ and forms the conventional superparaelectric state, in which the remanent polarization of the hysteresis loops disappears. The frozen superparaelectric state was observed by us previously in multiferroics– manganites $RMn_2O_5$(R = Gd or Bi) and $Gd_{0.8}Ce_{0.2}Mn_2O_5$[13–15], as well as $YCrO_3$[11]. The electric polarization in $YCrO_3$ was observed both along the c axis and in the transverse (001) plane. Temperatures $T_{fr}$ for these directions were much different and exceeded significantly $T_N$. In fact, it was established in [11] that there is only antiferroelectric ordering in the (001) plane in an ideal YCrO$_3$ crystal. The observed electric polarization was due to restricted polar domains formed near crystal defects of a certain type. Since it was stated in [1] that the ferroelectric ordering in RCrO$_3$ occurs only for magnetic R ions, we should analyze the electric polar properties in a RCrO$_3$ single crystal with a magnetic R ion. In this study, we perform for the first time a complex investigation of the polar properties of ErCrO$_3$ (ECO) single crystals with the magnetic R ion. It was found that the origin and properties of the observed electric polarization in ECO do not qualitatively differ from those of the previously studied YCrO$_3$ compound [11].

## 2. OBJECTS AND METHODS OF STUDY

ECO single crystals were grown by the method of spontaneous crystallization in a solution-melt and characterized by the sp. gr. Pbnm. In contrast to YCrO$_3$, the ECO crystals were grown using solvent Bi$_2$O$_3$. They were shaped as plates 2–3 mm thick and 3–5 mm$^2$ in area. The developed planes are oriented perpendicular to either the caxis or the [110] directions. To measure the dielectric permittivity, conductivity, and electric polarization, we formed parallel-plate capacitors ~0.5–0.8 mm thick and 3–4 mm2 in area. The conductivity and capacitance were determined by a Good Will LCR-819 impedance meter in the temperature range of 5–370 K. The electric polarization was measured by two methods: thermoactivated pyrocurrent method and PUND (positive up negative down) hysteresis loop method [16]. In the first case, the polarization was measured by a Keithly 6514 electrometer upon heating the sample with a constant rate of temperature change after the preliminary cooling of the sample in polarizing electric field E. When using the PUND method, the dynamic polarization of restricted polar domains was measured as a response to the series of successively applied two positive and two negative pulses of electric field E [16]. We chose a version of the PUND method that is adapted for measuring restricted polar domains and was applied by us previously in [11, 13–15]. The PUND hysteresis loop method makes it possible to subtract the parasitic conductivity contribution from the polarization. A comparative analysis of the properties of polarizations due to restricted polar domains, which were measured by the pyrocurrent and PUND methods in GdMn$_2$O$_5$ and Gd$_{0.8}$Ce$_{0.2}$Mn$_2$O$_5$, was presented in [15]. Reciprocal space maps for the (004) and (220) Bragg ref lections, yielding information about the ECO structural features, were also studied.

## 3. EXPERIMENTAL RESULTS AND ANALYSIS

### 3.1. ECO Dielectric Permittivity and Conductivity

The temperature dependences of the permittivity ε' and conductivity σ in ECO along the c axis for some frequencies are shown in, respectively, Figs. 3a and 3b. Similar dependences for the [110] direction are given in Figs. 3c and 3d.

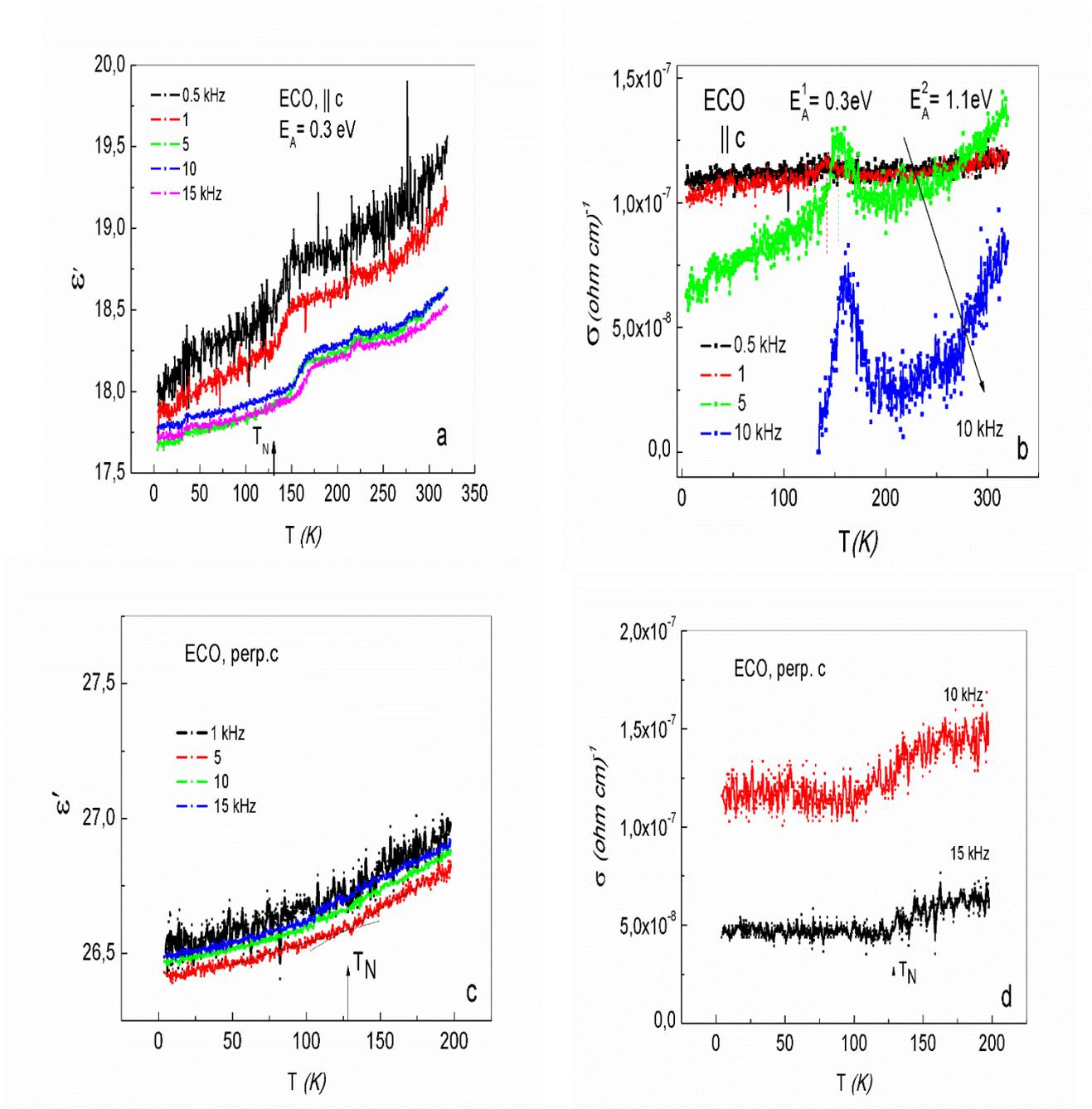

**Fig. 3.** Temperature dependences of the following parameters in ECO for some frequencies: (a) permittivity $\varepsilon'$ along the $c$ axis, (b) conductivity $\sigma$ along the $c$ axis, (c) $\varepsilon'$ along the [110] direction, and (d) $\sigma$ along the [110] direction.

Along the caxis, one can see relatively small permittivity $\varepsilon'$ weakly increasing with temperature and slightly decreasing with an increase in frequency. Frequency-dependent anomalies of $\varepsilon'$ in the form of steps are observed somewhat above $T_N$. The step temperatures increase with an increase in frequency and are described by the Arrhenius law with the activation barrier $E_A = 0.3$ eV (Fig. 3a).

Now let us consider the ECO conductivity. We deal with the real part of conductivity $\sigma_1 = \omega \varepsilon'' \varepsilon_0$ [17], which was calculated from the dielectric loss $\varepsilon''$ (the dissipation factor $\tan\delta = \varepsilon''/\varepsilon'$ was measured). Here, $\omega$ is the angular frequency and $\varepsilon_0 = \varepsilon'$ is the permittivity in a vacuum. Conductivity $\sigma_1$ (denoted as $\sigma$ below) depends on both frequency and temperature. The low-frequency part of conductivity is nondispersive and corresponds to percolation conductivity $\sigma_{dc}$ [17]. As can be seen in Fig. 3b, the conductivity $\sigma_{dc}$ in ECO along the c axis is low ($\sim 10^{-7}$ $(\Omega cm)^{-1}$) and changes only slightly with temperature at T< 330 K, which indicates fairly high dielectric characteristics of the ECO single-crystal matrix. Maxima of conductivity $\sigma$ (Fig. 3b) are observed at the temperatures of $\varepsilon'$ anomalies (Fig. 3a). The temperatures of conductivity maxima increase with an increase in frequency and are also described by the Arrhenius law with $E_A$= 0.3 eV. The presence of the frequency dispersions of $\varepsilon'$ and $\sigma$ anomalies does not make it possible to assign them to the ferroelectric phase transition. They are most likely indicative of the presence of restricted polar domains in the initial crystal matrix. However, the conductivity in these domains should increase with an increase in frequency [17], whereas in ECO (Fig. 3b) the conductivity decreases with an increase in frequency up to 250 K (although it exhibits the aforementioned maxima). Only at temperatures above 250 K, the conductivity at higher frequencies begins to exceed the low-frequency conductivity, that is, the local conductivity begins to manifest itself. The dependence of the temperatures of equality of percolation and local conductivities on frequency (Fig. 3b) obeys the Arrhenius law with the activation barrier $E_A$= 1.1 eV. Thus, along the c axis, two types of restricted polar domains with different activation barriers at their boundaries manifest themselves in ECO. The fact that the local conductivity of domains with small barriers at temperatures $kT \leq 0.3$ eV does not affect the percolation conductivity of the crystal matrix (Fig. 3b) indicates that these domains are not independent in the ECO crystal matrix and formed inside the domains with barriers of 1.1 eV. The proximity of the temperatures of $\varepsilon'$ and $\sigma$ anomalies to $T_N$ in ECO along the c axis is indicative of the influence of the magnetic ordering on restricted domains with barriers $E_A$= 0.3 eV. A comparison of the dielectric properties along the c axis for $YCrO_3$ [11] and ECO shows that $YCrO_3$ also contained restricted polar domains of two types; however, they were spatially separated in the crystal matrix. One of these types comprises restricted phase separation domains of magnetic origin with the activation barrier of 0.2 eV (existing from the lowest temperatures to T< 175 K), in which the high-frequency local conductivity exceeds the percolation conductivity beginning with the lowest temperatures. These domains were formed near impurity $Pb^{4+}$ ions, which substituted $Y^{3+}$ ions at certain crystal sites. The formation of these domains of magnetic origin was provided by the presence of pairs of chromium ions with different valences ($Cr^{3+}$ and $Cr^{2+}$) and electrons that recharge them [11]. Restricted polar domains of structural origin with the barriers $E_A = 0.49$ eV, the local conductivity of which manifested itself at T> 225 K, were formed simultaneously and independently near other lattice sites containing impurity $Pb^{2+}$ ions [11]. The above mentioned influence of magnetic ordering in ECO on polar structural domains along the c axis indicates that the restricted domains with $E_A = 0.3$ eV in ECO are most likely phase separation domains of magnetic origin containing ions with different valences ($Cr^{3+}$–$Cr^{2+}$). However, ECO contains no impurity $Pb^{4+}$ ions, which led to the formation of $Cr^{3+}$–$Cr^{2+}$ ion pairs in $YCrO_3$ due to the reaction $Y^{3+} = Pb^{4+} + e$, $Cr^{3+} + e = Cr^{2+}$. As was noted above, the ECO single crystals were grown by the method of spontaneous crystallization in a solution-melt using solvent $Bi_2O_3$. During the formation of ECO single crystals, $Bi^{3+}$ ions may substitute (with some probability) $Er^{3+}$ ions. Their ionic radii in the environment with a coordination number of 8 are 1.24 and 1.00 Å, respectively [18]. An X-ray fluorescence analysis showed that the content of $Bi^{3+}$ ions in ECO does not exceed 1.5%. $Bi^{3+}$ ions (as well as $Pb^{2+}$ ions) contain alone pairs of $6s^2$ electrons. It is known that the presence of $6s^2$ electrons leads to non-centrosymmetric distortion of the coordination environment of these ions [19]. As a result, impurity $Bi^{3+}$ ions in ECO violate the compensation of polarizations in the strongly correlated antiferroelectric matrix and form restricted polar domains of structural origin, to which we assign restricted domains with $E_A \approx 1.1$ eV.

………Let us discuss the possibility of occurrence of $Cr^{2+}$ and $Cr^{3+}$ ions in the domains of structural origin in ECO along the c axis. These ions most likely occur due to the tunneling of percolation-conductivity electrons of the ECO matrix through the barriers of polar structural domains. These electrons lead to the occurrence of $Cr^{2+}$ ions ($Cr^{3+} + e = Cr^{2+}$) and, eventually, phase separation domains with the barriers of 0.3 eV. The appearance of neighboring pairs of $Cr^{3+}$ and $Cr^{2+}$ ions with a finite probability of tunneling of $e_g$ electrons between these ions (double exchange [20]) stimulates an energetically favorable process of the formation of dynamically equilibrated phase separation domains. Ferromagnetic pairs of $Cr^{3+}$ and $Cr^{2+}$ and charge carriers ($e_g$ electrons that recharge these pairs) are accumulated in these domains, similarly as for $LaAMnO_3$ (A = Sr, Ca, or Ba) [21, 22] and in $RMn_2O_5$ [13–15]. Note that $Cr^{3+}$ ions are analogs of $Mn^{4+}$ ions (their 3d shells contain three $t_{2g}$ electrons in the triplet state and an unfilled orbital doublet). $Cr^{2+}$ ions are analogs of $Mn^{3+}$ ions and, along with three $t_{2g}$ electrons in the triplet state, contain one $e_g$ electron at the degenerate orbital doublet. The double exchange requires the presence of ferromagnetic spin moments Fz of $Cr^{3+}$– $Cr2^+$ ion pairs, which is implemented in ECO along the c axis at $T \leq T_N$. Since the phase separation domains are formed at the balance of strong interactions (double exchange, Jahn–Teller interaction, and Coulomb repulsion), they continue to exist in the paramagnetic state. However, the thermal activation of electrons from the phase separation domains is amplified at temperatures $kT \geq E_A = 0.3$ eV, which increases their concentration in the domains of structural origin. As a result, the number of Jahn–Teller $Cr^{2+}$ ions in octahedra increases in these domains. Deformation of these octahedra increases the dielectric permittivity (Fig. 3a). An increase in the electron concentration in the domains of structural origin also leads to the dominance of local conductivity of these domains at temperatures above 250 K (Fig. 3b).

In the [110] direction, the ECO permittivity is somewhat higher but depends more weakly on frequency and temperature (Fig. 3c). Slight changes in the slopes are observed in the temperature dependences of ε' near $T_N$ (Fig. 3c). The value of conductivity in this direction (Fig. 3d) is close to that observed in ECO along the c axis (Fig. 3b). The conductivity decreases with an increase in frequency, which also indicates the dominance of percolation conductivity [17]. Small jumps are observed in the temperature dependences of σ near $T_N$ (Fig. 3d). The proximity of ε' and σ anomalies in the [110] direction to $T_N$ is also indicative of some influence of the magnetic ordering in ECO on the properties of restricted polar domains. It is reasonable to relate this fact to the tunneling of percolation-conductivity matrix electrons through the high barriers of the domains of structural origin, which form Jahn–Teller Cr2+ ions in polar structural domains. The concentration of these electrons increases at T> TN, when eg electrons, which recharge Cr2+ –Cr3+ ion pairs in the phase separation domains along the c axis, are thermally activated, thus increasing the percolation conductivity. This leads to a rapid increase in ε' and jumps of σ (Figs. 3c, 3d) in the [110] direction as well at T> $T_N$.

Thus, we relate the observed anomalies of the dielectric properties of ECO in both directions to the presence of restricted polar domains of structural origin formed near impurity $Bi^{3+}$ ions, which substitute $Er^{3+}$ ions during the growth of single crystals. The influence of the ECO magnetic ordering near $T_N$ on the dielectric properties is due to the formation of phase separation domains at the tunneling of percolation-conductivity electrons of the crystal matrix to the restricted domains of structural origin.

Note that the comparison of the dielectric properties of ECO and $YCrO_3$ [11] shows that these proper-ties are similar in the [110] direction. They are due to restricted polar domains of structural origin, which are formed near strongly polarizing $Bi^{3+}$ (in ECO) and $Pb^{2+}$ (in $YCrO_3$) impurities. In this direction, relatively low permittivities and conductivities, which weakly depends on temperature and frequency, are characteristic of ECO and $YCrO_3$. This is caused by the existence of rigid antiferroelectric ordering in both crystals. The ECO dielectric properties along the c axis are almost similar to those in the [110] direction, whereas the dielectric properties in $YCrO_3$ along the c axis are significantly different. Much higher permittivities and conductivities along the c axis are observed in $YCrO_3$, which strongly depend on temperature and frequency [11]. We relate this difference to the role of $Er^{3+}$ ions in ECO, which are rigidly fixed along the c axis by the strong spin–orbit interaction. The rigidity of the ECO dielectric properties along the c axis and in the plane oriented perpendicular to the latter is due to, respectively, $Er^{3+}$ ions and antiferroelectric ordering. There is only antiferroelectric ordering in the (001) plane in $YCrO_3$.

3.2. Electric Polarization of ECO Measured by the Pyrocurrent Method

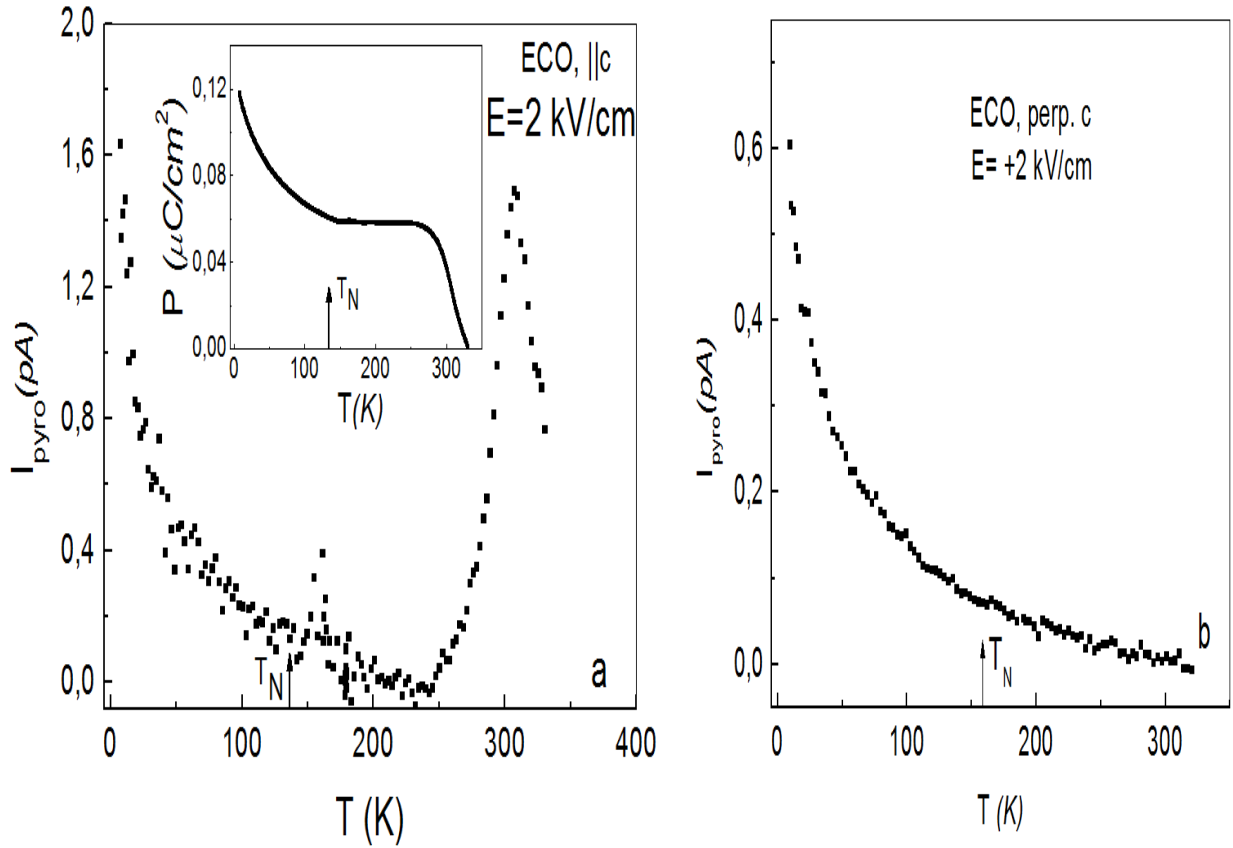

Fig. 4. Temperature dependences of the (a) pyroelectric current and electric polarization (in the inset) for ECO along the c axis and (b) pyroelectric current in the [110] direction.

Figure 4a shows the temperature dependence of the thermoactivated pyrocurrent for ECO along the c axis after cooling the sample in the applied field of E=2 kV/cm. The temperature dependence of the ECO polarization along the c axis is given in the inset. The main pyrocurrent maximum is observed at T= 310 K, which is much higher than $T_N$. A small additional maximum is observed at temperatures of 160–170 K, near which the ε' and σ anomalies were observed (Figs. 3a, 3b). In the low-temperature region, a sharp increase in the pyroelectric current is observed when approaching the extremely low temperature in our experiment (T≈ 8 K), which is most likely related to the appearance of the pyrocurrent maximum at the magnetic-ordering temperature of Er3+ ions (TN1=5K) along the c axis. Notably, the ferroelectric ordering occurs, which is induced by the magnetic ordering of rare-earth ions in orthoferrites and orthochromites; it was considered theoretically in [23]. At T> 8 K, one can see the temperature-delayed tail of the pyroelectric current (and, correspondingly, polarization) up to a small pyroelectric-current maximum in the vicinity of 160–170 K. We relate this maximum to the validity of the condition $kT \approx E_A = 0.3$ eV for the phase separation domains of magnetic origin. The polarization mainly caused by the polar domains of structural origin (~0.06 μC/cm$^2$) manifests itself at temperatures from 170 K to Tfr~ 350–360 K (an extrapolated value of the high-temperature slope of the pyrocurrent maximum at T= 310 K to Ipyro= 0 (Fig. 4a). In the [110] direction, the low-temperature behavior of the pyroelectric current is similar to that for the c axis, and no high-temperature maxima are observed up to 330 K (Fig. 4b), which does not make it possible to determine the polarization value in this direction by the pyrocurrent method.

## 3.3. Electric Polarization of ECO Measured by the PUND Hysteresis Loop Method

Figure 5a shows the hysteresis loops of electric polarization for ECO along the c axis for some temperatures. It can be seen that the maximum polarization for these loops ( ≈ 0.034 μC/cm$^2$) coincides with the remanent polarization (the loops are saturated). The temperature dependences of the remanent polarizations are given in Fig. 5b. A slight increase in the polarization begins at temperatures above $T_N$, which is consistent with the increase in the permittivity ε' (Fig. 3a). This can be related to above considered increase in the concentration of Jahn–Teller $Cr^{2+}$ ions in the restricted domains of structural origin with an increase in the probability of tunneling of thermally activated electrons through the barrier at the structural-domain boundaries. Deformation of octahedra with Cr2+ ions increases the polarization in polar restricted domains. At T≈ 200 K, the $P_c^{rem}$ value was measured by the PUND hysteresis loop method to be 0.036 μC/cm$^2$ (Fig. 5a). At the same time, the polarization value measured by the thermoactivated pyrocurrent method at the same temperature is 0.06μC/cm$^2$ (Fig. 4a). As was noted above, when the polarization is measured by the pyrocurrent method, it is amplified by the conductivity; the contribution from the latter to the polarization is subtracted in measurements by the PUND hysteresis loop method [13–15]. It was shown in [15] that, when measuring the polarization of restricted polar domains in the frozen superparaelectric state, the temperatures Tfr, below which this state exists, coincide for the measurements by the pyrocurrent and hysteresis loop methods and correspond to the condition $kT_{fr} \approx E_A$ at the boundaries of restricted polar domains. As was noted above, when the ECO polarization is measured by the pyrocurrent method along the c axis, the temperature $T_{fr}$ is ~350–360 K. Measurements of the polarization by the PUND method at the field orientation E|| [110] were carried out in the temperature range of 5–370 K. In the entire range, the polarization value was larger by a factor of 3 as compared with the c axis (compare Figs. 5a and 5c).

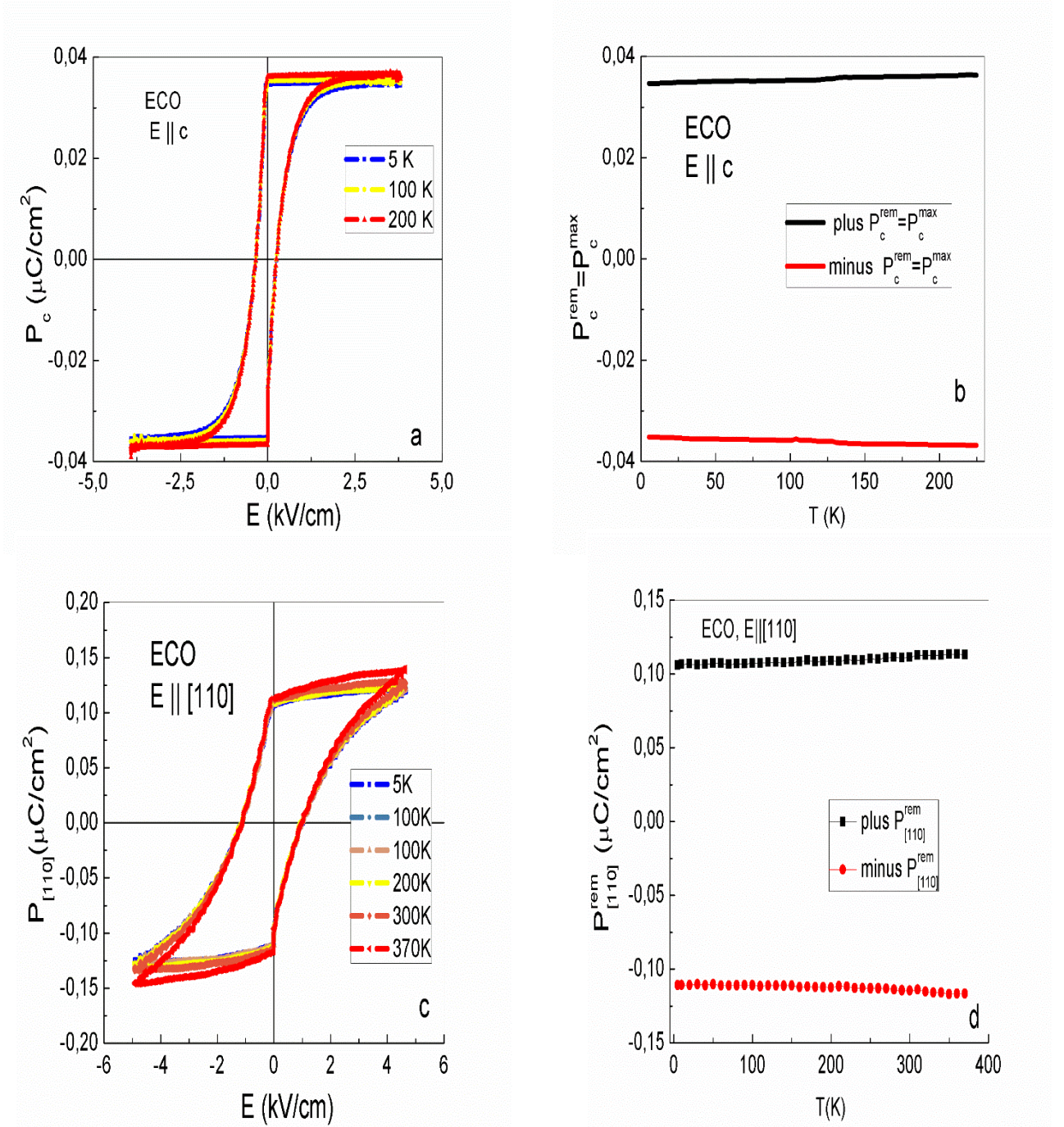

Fig. 5. (a, c) Electric polarization loops of ECO for the characteristic fixed temperatures and (b, d) temperature dependence of the remanent polarizations; (a, b) along the c axis and (c, d) in the [110] direction

The remanent polarization along the [110] direction slightly increases from 0.104μC/cm$^2$ at T< $T_N$ to 0.115 μC/cm$^2$ at T= 370 K (Fig. 5d). We assume that the restricted polar domains of structural origin formed near impurity $Bi^{3+}$ ions violate the compensation of the antiferroelectric-ordering sublattice polarizations in the [110] direction more significantly in comparison with the c axis. A smooth increase in the remanent polarization at T> $T_N$ in this direction is due to the same processes as for the c axis. The temperature $T_{fr}$ for the restricted polar domains of structural origin in the [110] direction exceeds 370 K.

## 4. HIGH-RESOLUTION X-RAY DIFFRACTION OF ECO

To confirm the presence of restricted polar domains in the ECO single-crystal matrix, we studied reciprocal space maps [24] for the (004) and (220) Bragg reflections (Figs. 6a and 6b, respectively). High-resolution triple-crystal X-ray diffractometry was applied with Ge crystals as a monochromator and analyzer in the (004) ref lection for CuKα1 radiation.

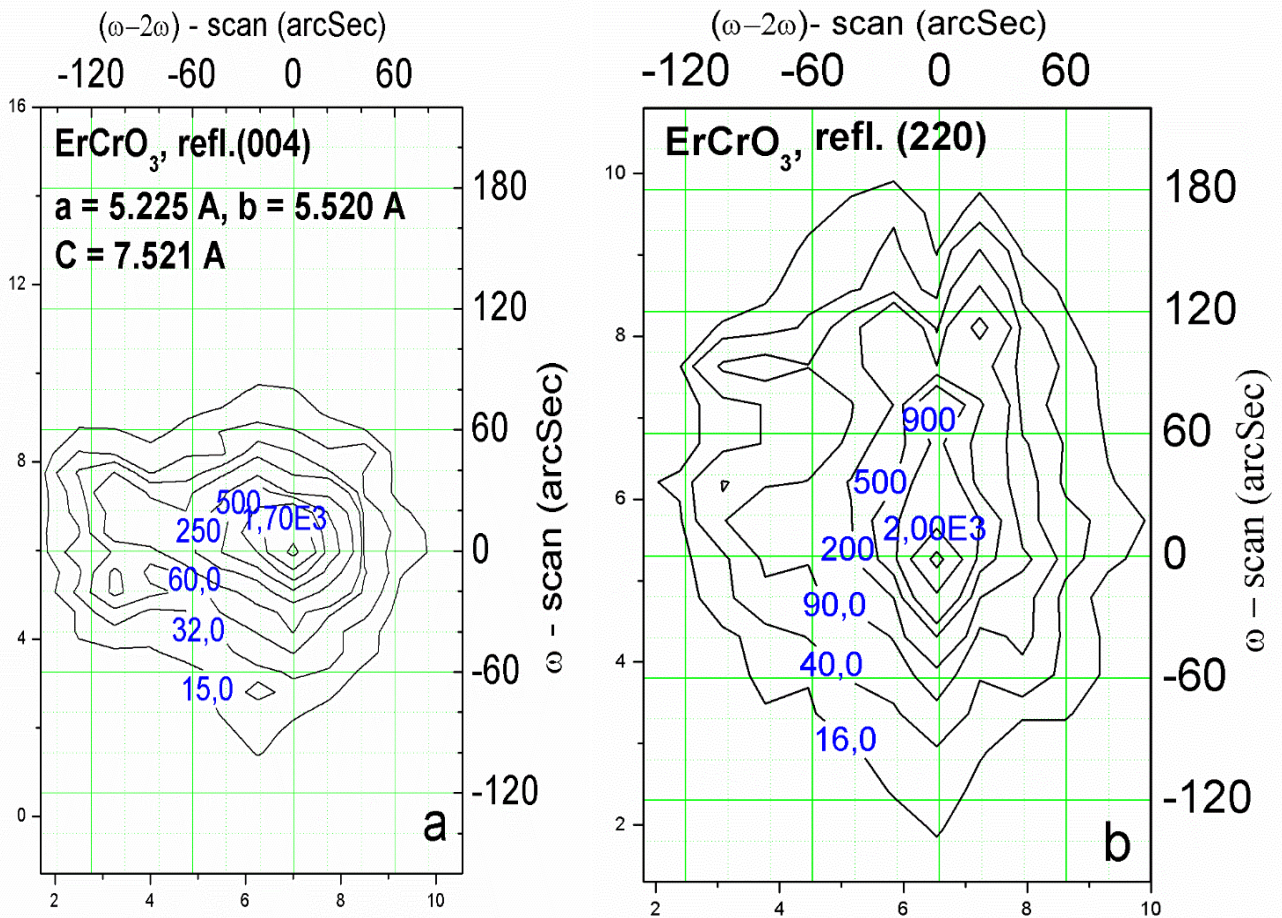

Fig. 6. Reciprocal space maps for the (a) (004) and (b) (220) Bragg reflections ECO.

The maps were recorded from natural mutually perpendicular sample faces in the [001] and [110] directions at room temperature. A comparison of the obtained contours of equal intensity for the (004) and (220) reflections reveals the differences between them. Approximately identical widths of the intensity distribution regions are observed for both reflections in the (ω–2ω) scan direction, whereas in the ω scan direction the (220) reflection exhibits a larger broadening of the region in comparison with (004). Remember that he broadening of (ω–2ω)-scan is caused by the distribution of the interplanar distances and the broadening of ω-scan is caused by mosaicity (misorientation of structure fragments) [24]. Taking this into account we can conclude that we deal with a fragmentary structure, in which misorientations in the directions perpendicular to the c axes dominate in comparison with the directions along the c axis. Notably, the contribution from the broadening due to the change in the lattice parameter is the same in both directions. The above analysis of the dielectric properties and electric polarization of real ECO crystals shows that there are restricted polar domains in the crystal matrix, which are formed near impurity $Bi^{3+}$ ions. Additional structural distortions (in comparison with the crystal matrix) appear in these domains. We assume that these restricted domains are responsible for the fragmentation of structural formations in the reciprocal space maps, which are observed in ECO (Figs. 6a, 6b). Note that the observed anisotropy of the values of ECO electric polarization (Figs. 5a–5d) corresponds to that of the observed misorientations in these maps.

## 5. CONCLUSIONS

Thus, uniform ferroelectric ordering in the temperature range of 5–370 K was not found in ECO single crystals in any of the crystal directions. However, the electric polarization induced by restricted polar domains, which were formed near impurity $Bi^{3+}$ ions substituting Er3+ ions during the single-crystal growth, was observed. These restricted polar domains of structural origin form the superparaelectric state. At sufficiently low temperatures, i.e., when kT is smaller than the activation barriers at the boundaries of these domains ($E_A$≈ 1.1 eV), they exhibit electric polarization (local ferroelectric ordering). At $kT_{fr} \leq E_A$, these domains form the frozen superparaelectric state, at which the domains do not undergo spontaneous repolarization. This state, theoretically considered in [12], should exhibit hysteresis loops with remanent polarization, which was observed in our experiments. The polarity of restricted domains is caused by the violation of compensation of the antiferroelectric state due to structural distortions in the domains. At temperatures $kT_{fr} \geq E_A$, the domains undergo spontaneous repolarization that is characteristic of the conventional superparaelectric state, at which the remanent polarization in the hysteresis loops disappears. The performed analysis of the experimental data showed that the conditions required for the existence of the frozen superparaelectric state at $kT_{fr}< E_A$ are satisfied in ECO single crystals for the observed restricted polar domains of structural origin. The temperatures $T_{fr}$ depend on the direction of the crystal axes and exceed significantly the magnetic-ordering temperature $T_N$. It was found that the restricted phase separation domains of magnetic origin affect only slightly the restricted-domain polarization up to temperatures that exceed $T_N$ by 30–40 K. This influence in the ECO crystal with $Er^{3+}$ magnetic ions was much weaker in comparison with $YCrO_3$ [11].

ACKNOWLEDGMENTS

This study were supported by the Russian Foundation for Basic Research, project no. 18-32-00241, and the Program No 1.4 "Actual Problems of Low-Temperature Physics" of the Presidium of the Russian Academy of Sciences.